# Multi-objective Optimisation Framework for Blue-Green Infrastructure Placement Using Detailed Flood Model

Asid Ur Rehman [1*], Vassilis Glenis [1], Elizabeth Lewis [2], Chris Kilsby [1]

[1] School of Engineering, Newcastle University, Newcastle Upon Tyne, NE1 7RU, UK

[2] School of Engineering, The University of Manchester, Manchester, M1 7HL, UK

[*] Correspondence: asid-ur-rehman2@newcastle.ac.uk

## Highlights

- Introducing a cost-benefit optimisation framework for Blue-Green Infrastructure.
- The use of a detailed flood model provides exact locations for interventions.
- Smaller-sized permeable interventions in many places offer better cost-benefits.

## Abstract

Designing city-scale Blue-Green Infrastructure (BGI) for flood risk management requires detailed and robust methods. This is due to the complex interaction of flow pathways and the need to assess cost-benefit trade-offs for various BGI options. This study aims to find a cost-effective Blue-Green Infrastructure placement scheme by developing an improved approach called the Cost Optimization Framework for Implementing blue-Green infrastructure (CONFIGURE). The optimisation framework integrates a detailed hydrodynamic flood simulation model with a multi-objective optimisation algorithm (Non-dominated Sorting Genetic Algorithm II). The use of a high-resolution flood simulation model ensures the explicit representation of BGI and other land use features to simulate flow pathways and surface flood risk accurately, while the optimisation algorithm guarantees achieving the best cost-benefit trade-offs for given BGI options. The current study uses the advanced CityCAT hydrodynamic flood model to evaluate the efficiency of the optimisation framework and the impact of location and size of permeable interventions on the optimisation process and subsequent cost-benefit trade-offs. This is achieved by dividing permeable surface areas into intervention zones of varying size and quantity. Furthermore, rainstorm events with 100-year and 30-year return periods are analysed to identify any common optimal solutions for different rainfall intensities. Depending on the number of intervention locations, the automated framework reliably achieves optimal BGI implementation solutions in a fraction of the time required to find the best solutions by trialling all possible options. Designing and optimising interventions with smaller sizes but many permeable zones saves a good fraction of investment. However, such a design scheme requires more computational time to find optimal options. Furthermore, the optimal spatial configuration of BGI varies with different rainstorm severities, suggesting a need for careful selection of the rainstorm return period. Based on the results, CONFIGURE shows promise in devising sustainable urban flood risk management designs.

**Graphical abstract**

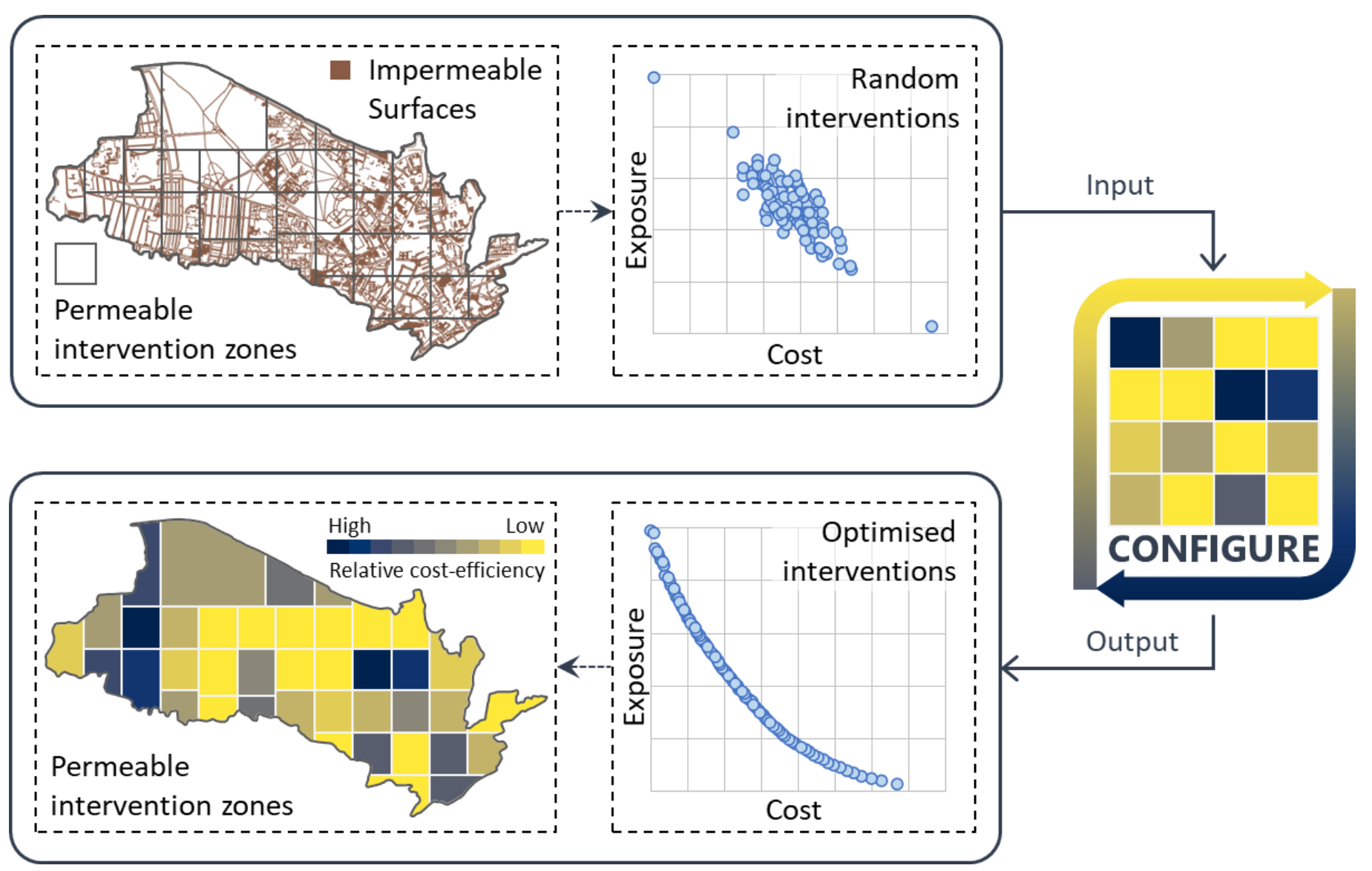

## 1 Introduction

The projected increase in the frequency and severity of rainfall (Kendon et al., 2023; Robinson et al., 2021) combined with ongoing urbanisation (Miller & Hutchins, 2017) make cities more prone to flash flooding. Conventional grey infrastructure-based stormwater management approaches, such as underground drainage and combined sewer pipe networks, have demonstrated limitations in conveying surface runoff during extreme weather events because they are designed to handle only a certain amount of rainfall (Abduljaleel & Demissie, 2021; POST, 2007). Additionally, combined sewer overflows (CSOs) cause environmental pollution and pose risks to public health (Botturi et al., 2021). Moreover, all these highly engineered approaches incur high financial and environmental costs (Rosenbloom, 2018). Therefore, new and sustainable approaches are required to manage floods in the catchment and cities efficiently (POST, 2007). Blue-Green Infrastructure (BGI) features or Low-Impact Development (LID) such as permeable surfaces, detention ponds, green roofs, rain gardens, swales, bioretention cells, and water butts offer a promising solution for sustainable urban flood risk management (O'Donnell et al., 2020). In contrast to grey infrastructure, BGI follows the concept of 'managing flood at its source' i.e., designing natural or semi-natural interventions to reduce the surface run-off volume and intensity by mimicking natural hydrological processes of infiltration, evaporation, interception, and storage (Ahiablame et al., 2012). Thus, BGI directly reduces the pressure on existing urban stormwater management systems. In addition to their potential to mitigate urban flash flooding, these semi-natural features have other multi-functional benefits such as water conservation, water quality improvement, biodiversity enhancement, air quality improvement, and urban heat island effect mitigation (Rodriguez et al., 2021). When it comes to practical implementation, there has been limited work on comprehensively assessing the cost-benefits of Blue-Green Infrastructure (BGI) due to the inclusion of many non-monetary values (Chen et al., 2020). A recent study by Wang, J., & Banzhaf, E. (2018) highlighted gaps between Green Infrastructure (GI) mapping and GI functional analysis, making it difficult to evaluate the actual impacts of GI. To address this challenge, researchers such as Choi et al. (2021) and Gordon et al. (2018) have developed detailed frameworks to assess both monetary and non-monetary benefits, including environmental, social, technical, and economic performance metrics. Recently, Chen et al. (2020) evaluated selected ecosystem services provided by urban green infrastructure (UGI) and they converted those services into monetary values as well. However, they did not consider the costs associated with UGI implementation. Earlier, Beauchamp and Adamowski (2012) assessed the cost-benefits of GI for a housing development project and found them to be approximately equal to those of conventional grey infrastructure. By factoring in other environmental benefits, the authors suggested adopting a GI approach for new housing developments. In line with these recommendations, local government authorities have already started incorporating BGI into their strategic surface water management plans (Wheeler, 2016).

Despite having numerous advantages, cost-effective deployment of BGI is a big challenge. The effectiveness of BGI strongly relies on their spatial configuration i.e., features type, size, and their location of deployment (Huang et al., 2022; Perez-Pedini et al., 2005). These configuration parameters are often evaluated in a hydrodynamic flood model to find their optimal settings (D'Ambrosio et al., 2022; Rodriguez et al., 2021). However, when there is a wide range of configuration parameter values and combinations, their testing poses a significant computational challenge, even with modern computing systems. To overcome this challenge, researchers have adopted different statistical and analytical methods, which include simple scenario-based analysis (Abduljaleel & Demissie, 2021; D'Ambrosio et al., 2022; Webber et al., 2020), flood source-receptor-based scenarios (Vercruysse et al., 2019), multi-criteria and analytical hierarchy approaches (Alves et al., 2018; Joshi et al., 2021; L. Li et al., 2020), and exploratory spatial data analysis (ESDA) (Rodriguez et al., 2021). While these methods seem quite effective in deriving time-efficient solutions, they also fall short of providing insights into whether the identified solutions are the ultimate cost-effective choices. To address the challenge of deriving the most cost-effective solutions, researchers have adapted Evolutionary Algorithms (EAs) or Multi-objective Optimisation Algorithms (MOOAs), which are commonly used for multi-objective optimisations (Lu et al., 2022; Maier et al., 2019; Seyedashraf et al., 2021).

The Storm Water Management Model (SWMM) (Rossman & others, 2010) is the tool most integrated with MOOAs (Zhang & Jia, 2023) to assess the efficiency of Blue-Green Infrastructure (BGI) in reducing peak-flows and/or total volume in underground drainage systems during a rainstorm event. For example, Wang et al., (2023) utilised future climate scenarios of rainfall and a combination of graph theory and genetic algorithm to optimise spatial green-grey layouts, employing the SWMM model. Similarly, Yao et al., (2022) maximised monetised net benefits against the cost of coupled green-grey infrastructure for different return periods by integrating the Non-dominated Sorting Genetic Algorithm

II (NSGA-II) with SWMM. Various other studies (Gao et al., 2022; Hassani et al., 2023; S. Li et al., 2022; Lu et al., 2022; Nazari et al., 2023; Rezaei et al., 2021; J. Wang et al., 2022; Zhu et al., 2023) have also integrated hydrodynamic models, predominantly SWMM, with various optimisation algorithms, including Genetic Algorithm (GA), Particle Swarm Optimisation (PSO), and Generalised Differential Evolution (GDE3), to devise cost-effective low-impact development or green-grey infrastructure designs for rainstorm water management. Despite its wide application, SWMM has some limitations. As a semi-distributed model, it treats each sub-catchment as a hydrological response unit and estimates surface run-off without providing any overland flow routing among the sub-catchments. Consequently, SWMM cannot incorporate over-surface land use features, such as buildings and roads, exclusively, nor can it simulate 2D surface water flows, such as pluvial and/or fluvial flooding. These limitations are overcome by using fully distributed hydrodynamic flood models such as CityCAT (Glenis et al., 2018), TELEMAC (Hervouet, 1999), and InfoWorks ICM (Innovyze, 2013). These models explicitly incorporate land use features and their characteristics on a high-spatial-resolution computational grid and normally apply 2D shallow water equations to simulate 2D surface flows and/or 1D drainage flows (Pina et al., 2016). During the simulation, they capture the dynamic interaction of over-surface features with water flow paths to provide accurate floodwater depths and velocities.

In contrast to their use in detailed flood risk assessments (Pregnolato et al., 2016; Sun et al., 2021), the application of 2D surface models in MOOA-based BGI design remains rare. One potential factor is the need for detailed input data to set up these models and the considerable computational resources required to run them (Hill et al., 2023). However, the continuous production of accurate, high-scale datasets and advancements in computational technology have reduced the magnitude of this challenge. Another possible explanation is the lack of readily adaptable automated optimisation tools for such types of models. Although existing tools like *Pymoo* (Blank & Deb, 2020) and *DEAP* (Fortin et al., 2012) are open access, their universal nature may present challenges when integrating them with detailed hydrodynamic models. Users may need to invest considerable time and effort to understand the problem formulation process of these tools to frame their optimisation problem accurately and efficiently. Additionally, the evolutionary operations in multi-objective algorithms can often be sensitive to the nature and type of optimisation problems being addressed (Karafotias et al., 2015). Therefore, fine-tuning evolutionary operators in generic optimisation tools can be a challenging task, especially for inexperienced users. These reviewed challenges highlight two research gaps: (1) the need for the development of an easily adaptable optimisation tool tailored specifically for BGI design, and (2) the integration of such a tool with a fully distributed hydrodynamic flood model to optimise BGI placement by explicitly representing BGI features in the model to accurately simulate 2D surface flows and assess surface flood risks.

This study aims to address the identified research gaps by developing a multi-objective optimisation framework and integrating it with a detailed hydrodynamic flood model. The novelty of the current research lies in precisely locating the optimal sites for Blue-Green Infrastructure, such as permeable surfaces, using a detailed hydrodynamic model for optimization. The specific objectives of this research work are: (i) creating a multi-objective optimisation framework by incorporating the requirements of BGI features and integrating this framework with a detailed hydrodynamic model, and (ii) demonstrating the functionality and efficiency of the newly developed optimization framework for determining the locations, sizes, and spatial configurations of permeable surfaces. The remaining part of this paper is structured as follows: selection of an optimisation algorithm and detailed functionality of the optimisation framework including its application is presented in Section 2, the results, and discussions along with study limitations and future recommendations are presented in Section 3, and finally, concluding remarks are given in Section 4.

## 2 Material and methods

### 2.1 Selection of a MOOA

MOOAs normally tackle optimisation problems by iteratively refining a population of candidate solutions, based on objective functions, decision variables, and evolutionary processes, to find a new population (Venter, 2010). The iterative process continues until a termination criterion is met to achieve a set of best or near-best solutions, known as Pareto optimal solutions, that represent the best trade-offs between conflicting objectives (Maier et al., 2019). Optimisation algorithms vary widely, and no single algorithm universally outperforms others; the choice depends on factors such as the problem type and optimisation parameters (Z. Wang et al., 2022). One common approach to finding the most effective one is implementing multiple optimisation algorithms simultaneously (Yao et al., 2022). However, this approach is impractical with high-resolution hydrodynamic models due to their high computational demand. Instead, this study assessed different MOOAs based on their functional characteristics to select a viable option. Key traits defining an algorithm's effectiveness include its exploration and exploitation capabilities (Lin & Gen, 2009). Exploration represents an algorithm's ability to discover a wide range of unique solutions to find the best one, while exploitation refers to focusing the search on a specific direction to achieve optimal solutions quickly. Genetic Algorithm (GA) (Holland, 1992) and Differential Evolution (DE) (Storn & Price, 1997) prioritise exploration by maintaining diverse solution populations. In contrast, Particle Swarm Optimisation (PSO) (Kennedy & Eberhart, 1995) and Simulated Annealing (SA) (Kirkpatrick et al., 1983) focus on exploitation. PSO exploits swarm knowledge, while SA implements a cooling schedule to optimise within promising regions. When considered individually, the probabilistic nature of SA can aid in bypassing local optima, but it may encounter slow convergence (Ram et al., 1996), especially in high-dimensional problems. DE performs well for continuous problems but may face challenges in discrete or combinatorial optimisation (Slowik & Kwasnicka, 2020). Despite its efficiency and wide application, PSO is vulnerable to being trapped in local optima (Couceiro et al., 2016). In contrast, GA demonstrates excellent exploration capabilities, allowing it to reach global optimal solutions while offering a simple representation of candidate solutions suitable for both continuous and discrete optimisation problems in varying dimensional complexities. GA's main constraint is its low convergence speed, which can be mitigated by using its natural parallelisation support (Roberge et al., 2013; Verma et al., 2021). Considering these observations and anticipating the most suitable way of formulating the BGI optimisation problem, the enhanced multi-objective version of GA, known as Non-dominated Sorting Genetic Algorithm II (NSGA-II), was selected for the proposed optimisation framework. Although several NSGA-II variants exist, such as Epsilon-NSGA-II (Kollat & Reed, 2006) and Dynamic-NSGA-II (Deb et al., 2007), the original version of NSGA-II was selected because it is simple, well-understood, easy to implement, and offers good performance. NSGA-II utilises non-dominated sorting to select the best solutions and employs crowding distance techniques to maintain population diversity, facilitating better exploration in the search space (Deb et al., 2002).

## 2.2 Designing of BGI optimisation framework

The proposed BGI optimisation framework is named *CONFIGURE*, which stands for *Cost OptimisatioN Framework for Implementing blue-Green infrastructURE*. Figure 1 presents a generalised schematic diagram depicting the functionality of CONFIGURE. The entire framework is coded in Python and the script is freely available (see Appendix C: Availability of Python code) for integration with any hydrodynamic flood model of the user's choice. The main highlight of the CONFIGURE code is the implementation of Python's parallel processing functionality, allowing multiple instances of the hydrodynamic model to run simultaneously and thereby reducing convergence time.

In terms of functionality, CONFIGURE consists of five key components (Figure 1): problem framing, initial fitness evaluation, genetic operations, derivation of new generations, and optimised output. For a detailed overview of multi-objective optimization, readers can refer to Maier et al. (2019). Subsequent subsections will briefly discuss these five components to understand CONFIGURE's core functionality.

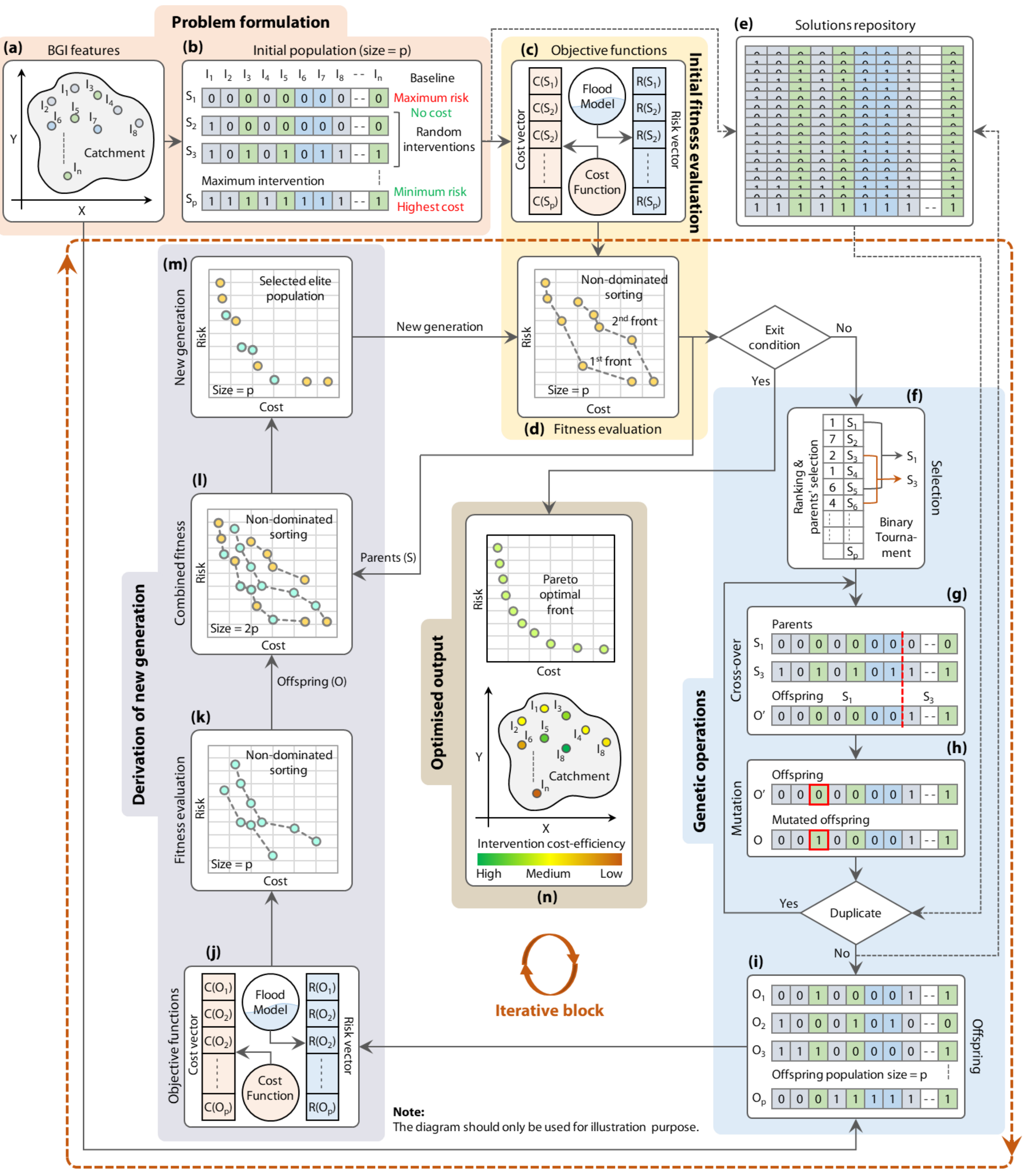

Figure 1. Generalised schematic diagram depicting the functionality of CONFIGURE at various stages: (a) BGI feature design, (b) solutions coding for initial population, (c) initial population simulation, (d) fitness evaluation in objective space, (e) solutions repository, (f) parents' selection, (g) cross-over, (h) mutation, (i) offspring population, (j) offspring population simulation, (k) offspring fitness evaluation, (l) parent + offspring population, (m) selected elite population

### 2.2.1 Problem framing

***BGI design and representation in hydrodynamic modelling:*** The selection of BGI types for optimisation depends on their suitability for addressing flooding in specific areas. Green roofs, permeable surfaces, and rain gardens typically handle floodwater at its source, while detention ponds and swales generally intercept flood pathways (Ahiablame et al., 2012). For modelling, BGI features are usually generated in standard GIS formats, such as shapefiles or geodatabases, with key attributes including feature sizes, water storage capacity, infiltration properties, and friction coefficients. Once the input data is prepared, the fully distributed model creates a computational grid, either a regular or irregular mesh grid, and explicitly represents land use and BGI features on that grid. This process is hypothetically illustrated in Figure 2. Each cell of the grid (depicted by the red cell on the grid, for example) receives relevant inputs, such as amount of rainfall, elevation, and land use/BGI properties, and uses numerical equations to calculate the water depth specific to that cell. The model then calculates surface runoff by considering the elevations and water depths of neighbouring cells. Figure 2a depicts maximum water depths calculated without BGI interventions (baseline flood modelling), while Figure 2b represents water depths with BGI features implemented. Such flood depth maps are used to assess the levels of risk to infrastructure, properties and/or businesses before and after implementing BGI.

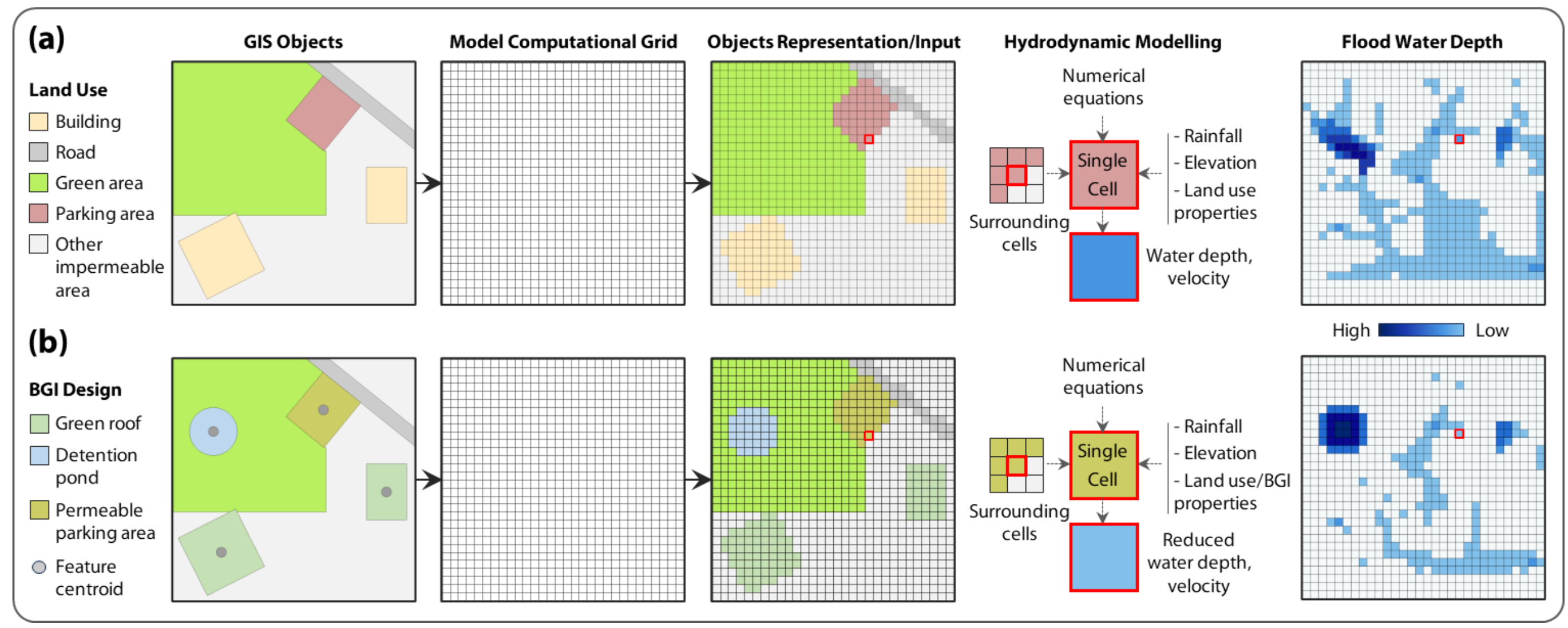

Figure 2. High-resolution hydrodynamic simulations (a) without and (b) with BGI feature representation.

***Decision variables and objective functions:*** Decision variables are the locations of BGI features, as shown in Figure 1, and represented in Equation (1)

$$I = (I_1, I_2, I_3, \ldots. I_n) \quad (1)$$

Each intervention location is assigned a unique index ($I_{j\ j=1,2,..n}$), and the decision space includes 'n' intervention locations of different BGI types, resulting in 'n' decision variables. CONFIGURE aims to minimise the lifecycle cost (C) of the BGI and the associated levels of risk (R). These objectives can be represented as follows (Seyedashraf et al., 2021):

$$Minmise: F(I) = (F_C, F_R) \quad (2)$$

The life cycle cost for the $j^{th}$ BGI feature can be calculated as follows:

$$C(I_j) = \left(c_c(I_j) + c_o(I_j)\right) \times s(I_j) \quad j = 1, \ldots, n \quad (3)$$

In Equation (3), $c_c(I_j)$ and $c_o(I_j)$ represent the unit size capital cost and unit size operational cost for the $j^{th}$ BGI feature respectively, while $s(I_j)$ is the size of the $j^{th}$ BGI feature. The capital cost refers to the one-time cost associated with installing the BGI features, whereas the operational cost represents the rolling expenses related to maintaining and managing the BGI features for a certain lifetime. The risk values (R) depend on the type of risk being considered.

***Candidate solution representation**:* The binary combinatorial method is used to represent locations of BGI features in the candidate solutions. As illustrated in Figure 1b, each bin of candidate solutions represents a unique BGI location and can be assigned a value of 1 (feature present) or 0 (feature not present). A set of 'p-2' candidate solutions is randomly generated, with the $1^{st}$ and $p^{th}$ candidate solutions designated as baseline (no intervention) and maximum (all intervention) scenarios, respectively. The total number of possible BGI candidate solutions can be calculated using $2^n$. Following binary coding, candidate solutions undergo initial fitness evaluation and are stored in a solution repository, as shown in Figure 1(c, e).

### 2.2.2 Initial fitness evaluation

Initial fitness evaluation is the process of measuring the fitness of candidate solutions by examining objective function values (cost and risk) for the initial population, as illustrated in Figure 1c. The purpose of the initial fitness evaluation is to identify the fittest parents, which in turn produce potentially superior offspring. The lifecycle cost ($F_C$) of the z[th] candidate solution ($S_z$) can be calculated using the following equation:

$$F_C = C(S_z) = \sum_{j=1}^{n} C(I_j) \quad \begin{matrix} where\ z = 1,2,3....p \\ I_j \neq 0 \end{matrix} \tag{4}$$

The risk function ($F_R$) is user-defined and is calculated from the outputs (flood depths) of the selected flood simulation model. Based on objective function values, a fitness function calculates fitness scores as shown in Figure 1d. The fitness of solutions is calculated by performing non-dominated sorting and crowding distance calculations (Deb et al., 2002), which leads to the emergence of various fronts, with the first front comprising superior solutions. After fitness evaluation, the algorithm checks the termination condition (further details are given in subsection 2.2.5). If the exit condition is met, the optimisation process terminates. Otherwise, the algorithm proceeds to the next stage.

### 2.2.3 Genetic operations

Genetic operations include parent selection, cross-over, and mutation to produce offspring solutions. CONFIGURE uses binary tournament selection (Figure 1f), random one-point cross-over (Figure 1g), and random single bit-flip mutation (Figure 1h) operators to create offspring. The incorporation of random positioning in cross-over and mutation operations enables the algorithm to explore the search space more effectively, thus mitigating premature convergence (Please refer to *Parameter sensitivity analysis in Supplementary Information S5* for further details). Additionally, the framework introduces a solution repository (Figure 1e) to store solutions generated across generations, thereby preventing duplication during offspring creation.

### 2.2.4 Derivation of the new generation

The new offspring population (as depicted in Figure 1i) is combined with the parent population to jointly calculate the overall fitness, as shown in Figure 1l. Through this controlled elitism, the new generation (Figure 1m) contains the best solutions from both the parent and offspring populations.

Subsequently, this new generation becomes the parents, and the entire process is repeated until a termination criterion is met.

### 2.2.5 Optimised output

CONFIGURE uses the generation count as a termination criterion to halt the optimisation process. The output of CONFIGURE comprises a set of solutions represented by the Pareto optimal front, as depicted in Figure 3a-b. The Pareto front offers a range of optimal BGI deployment options to end users. Additionally, CONFIGURE introduces an innovative spatial classification scheme (Figure 3c) for BGI features, based on their contribution to the Pareto front. BGI features that contribute most to optimal solutions are deemed highly cost-effective, and vice versa. This classification of BGI can significantly aid decision-makers in prioritising spatially for cost-effective BGI design. CONFIGURE Python code provides optimised outputs in comma-separated values (CSV) file format, including (1) all generations data, (2) optimal generation data, and (3) the contribution of BGI features to optimal solutions. The BGI contribution data can be linked to the shapefile of BGI features using their IDs to create a BGI contribution map (Figure 3c).

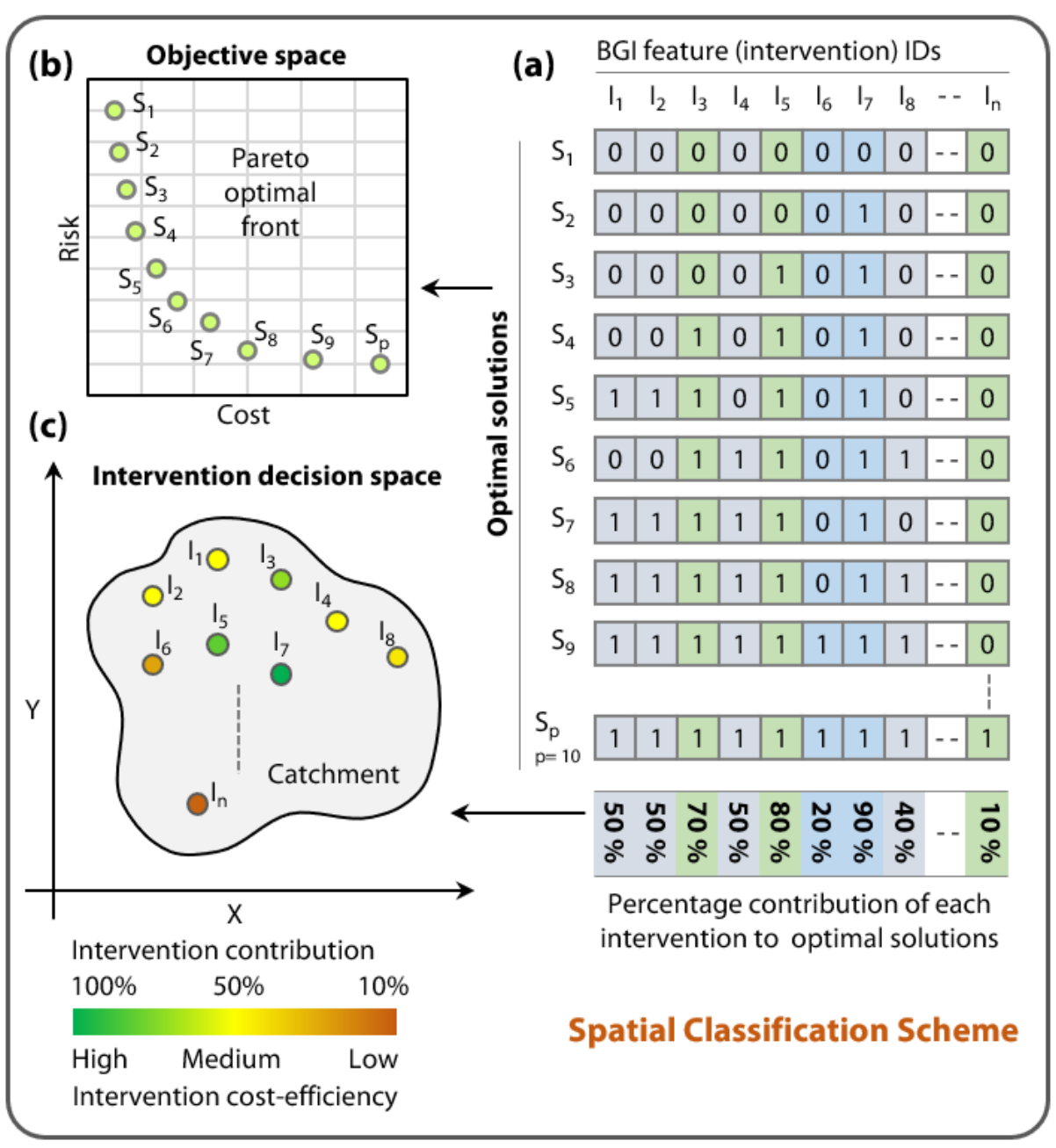

Figure 3. (a) optimal solutions, (b) Pareto front, (c) BGI intervention efficiency.

## 2.3 Case study

### 2.3.1 Study area

The study area shown in Figure 4 is the catchment of Newcastle upon Tyne's city centre. The catchment spans a total area of 5.3 $km^2$ which comprises 43.2% green space (including parks, moors, playgrounds, residential gardens, and roadside green belts), 32.5% impervious surfaces (such as roads, roadside pavements, and paths), and 24.3% buildings. The catchment's maximum elevation is 128 m, with a slope (northwest-southeast) of approximately 3.3%.

Newcastle has a history of both fluvial and pluvial floods since 1339 (Newcastle City Council, 2016). More recently, the city experienced a flash flood caused by an exceptional '*Thunder Thursday*' rainstorm on the 28th of June 2012. The city received 26 mm of rainfall in 30 minutes, 32 mm in 1 hour, and 49 mm in 2 hours (Environment Agency, 2012). *Thunder Thursday* was estimated to be around the 100-year rainstorm event, providing a strong basis for modelling the event and designing BGI interventions to apply and test the CONFIGURE framework.

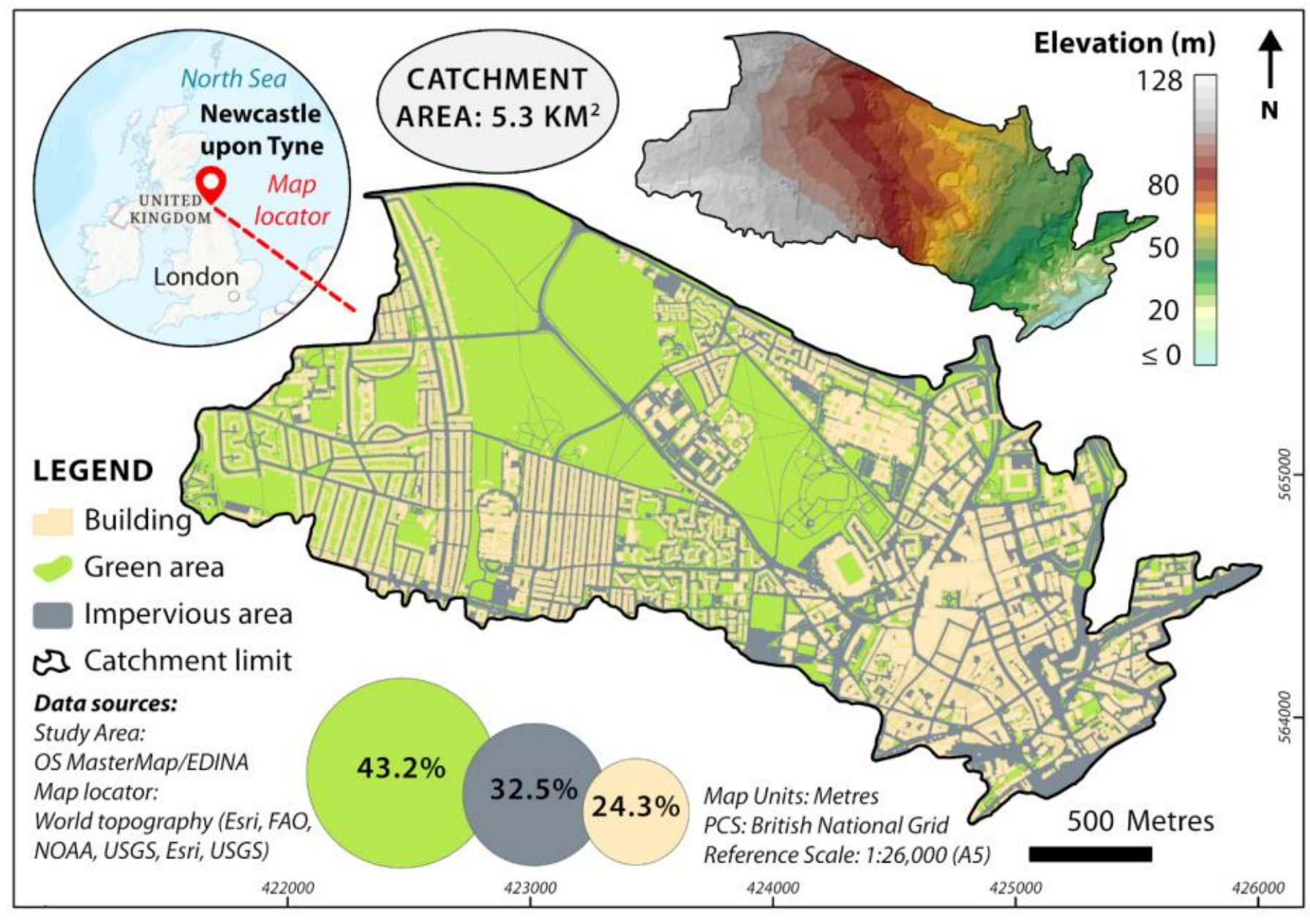

Figure 4. Map of the study area

### 2.3.2 Hydrodynamic modelling and exposure estimation method

The City Catchment Analysis Tool (CityCAT) was used to simulate flood hydrodynamics in the study area catchment. CityCAT, an advanced hydrodynamic model, is designed for high-resolution simulation and analysis of surface water flooding (Glenis et al., 2018). It operates in 2D and 1D/2D modes, offering a fully coupled approach. Moreover, the model is fully equipped to explicitly represent and simulate BGI features, enabling assessment of their potential in reducing urban flood risk (Kilsby et al., 2020).

CityCAT uses commonly available spatial datasets, including high-resolution Digital Elevation Models (DEMs) and geospatial vector datasets, to represent urban features such as buildings, green/permeable surfaces, and impermeable areas like roads and pavements. Figure 5(a, b) illustrates CityCAT's core functionality and outputs, beginning with the creation of a computational grid based on a DEM's spatial resolution. In urban areas, flow path accuracy is enhanced, and simulation time is reduced by excluding building footprints from the grid using building hole approach (Iliadis et al., 2023). The removed building cells are stored for later use in roof drainage algorithms (green roof interventions), while the rainfall that falls on the building cells is diverted to grid cells surrounding the building footprint. Each cell in the computational grid retrieves elevation data from the DEM and land use properties (surface friction, infiltration) from the land use vector dataset. Additional features like distributed rainfall, lakes, and ponds (with different friction coefficients, and soil properties) can also be associated with grid cells. Simulated rainfall can be linked with time-dependent boundary conditions of flow and/or water depth at spatial domain boundaries. Additionally, the model can efficiently integrate subsurface sewer networks for fully coupled surface-subsurface flood simulations, but this adds considerably to computation times.

The model estimates infiltration over permeable areas using the Green-Ampt method, considering soil properties such as hydraulic conductivity, porosity, and suction head. CityCAT's subsurface drainage component is based on the mathematical model for mixed flow in pipes presented by Bourdarias et al. (2012). The model uses the St. Venant equations and a conservative form of the equations for pressurised flow derived from the compressible Euler equations. Figure 5(b) displays the outputs of CityCAT, which include time series of flood depths, flow velocities, and volumes in and out of manholes and gully drains.

To represent Blue-Green Infrastructure, CityCAT includes built-in storage algorithms for interventions such as green roofs and rain barrels, based on user-defined storage capacities. Surface features like detention ponds and rivers are represented using Digital Elevation Models (DEM), while permeable features are modelled by assigning different infiltration and friction coefficient values.

CityCAT has undergone successful validation in various cities, including validation against analytic solutions of flows and laboratory datasets. Glenis et al. (2018) used an analytic solution and data from a physical model study of a dam break to demonstrate a strong agreement between the reference data and the CityCAT simulated numerical values of depth and velocity. Similarly, field validation extends to applications in Newcastle, where simulated flood depths for a real event in 2012 were checked using social media images (Kutija et al., 2014), validation of property flooding was conducted using local authority surveys (Bertsch et al., 2022) and a university campus survey (Iliadis et al., 2023). The results of these validation studies demonstrate a good agreement between the CityCAT-modelled flooded properties and the reference survey data.

**Exposure estimation:** The exposure calculation tool developed by Bertsch et al. (2022) assesses the number of buildings exposed to varying flood depths. By considering building characteristics, the tool evaluates exposure levels based on water depths surrounding the buildings. It achieves this by creating a spatial buffer around each building, as shown in Figure 5(c). The buffer's size is proportional to the dimensions of the water depth grid cells, and it intersects grid cells surrounding the buildings. The tool then extracts water depth information from these intersected grid cells and determines flood exposure levels for each building based on mean depth and $90^{th}$ percentile criteria from Figure 5(c).

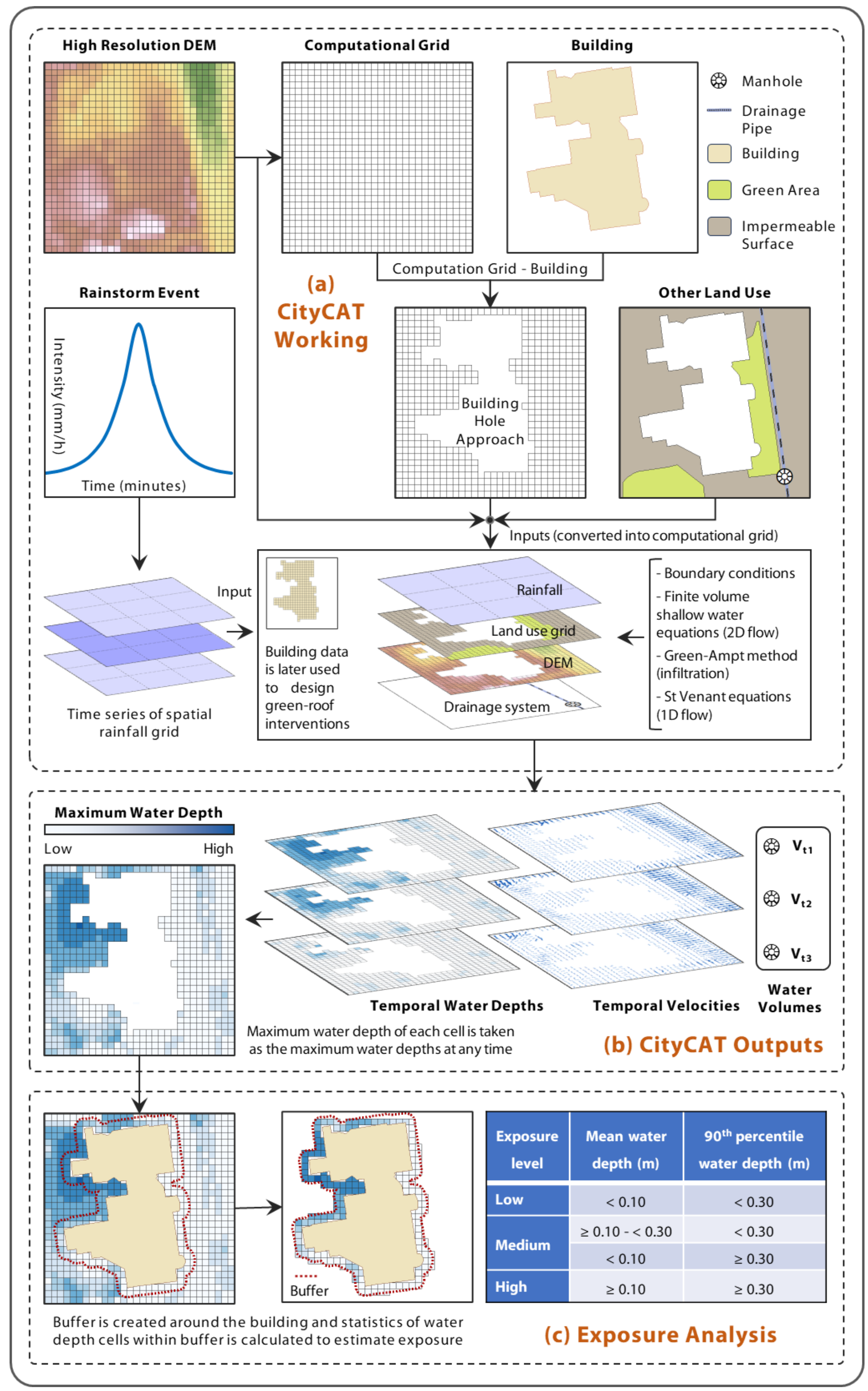

| Exposure level | Mean water depth (m) | 90th percentile water depth (m) |
| --- | --- | --- |
| Low | < 0.10 | < 0.30 |
| Medium | ≥ 0.10 - < 0.30 | < 0.30 |
| | < 0.10 | ≥ 0.30 |
| High | ≥ 0.10 | ≥ 0.30 |

Figure 5. (a) working principle of the CityCAT model and (b) its outputs (Adapted from Glenis et al., 2018), (c) building-level exposure calculation (Adapted from Bertsch et al., 2022).

### 2.3.3 Rainstorm design and CityCAT settings

Two synthetic rainfall events were generated using the depth-duration-frequency (DDF) model and storm profiling methods recommended by Faulkner (1999). These events have return periods of 100 and 30 years, each lasting half an hour. The following equation represents the DDF model for rainstorms with a duration of 12 hours or less:

$$lnR = (cy + d_1)\, lnD + ey + f \quad where\ y = -ln\left[-ln\left(1 - \frac{1}{T}\right)\right]$$

The variables in the DDF model are defined as follows: R is the rainfall depth, D is the duration, y is the Gumbel reduced variate, T is the return period, and c, $d_1$, e, f are catchment descriptors. The values for these descriptors can be obtained from the Flood Estimation Handbook (FEH) website (https://fehweb.ceh.ac.uk/). Using the DDF model, the rainfall amounts for the 100-year and 30-year return period events were calculated to be 31.1 mm and 21.9 mm, respectively. The temporal distribution profile of the rainstorms was then generated using the Flood Studies Report (FSR) method (Institute of Hydrology, 1975) for an urbanised catchment. The relevant formula is provided below.

$$y = \frac{1 - a^z}{1 - a}\ where\ z = x^b$$

Where y = proportion of the rainfall depth that falls within the proportion x of the total storm duration, centred on the peak. The parameters a and b have fixed values. The resulting rainstorm profiles are shown in *Supplementary Information S9*.

In this case study, only the 2D surface water simulation module of CityCAT was used to simulate surface runoff for the designed rainstorm events Details of the hydrodynamic parameters and their respective values used in CityCAT can be found in *Supplementary Information S1*.

### 2.3.4 BGI intervention design

Permeable flood risk management interventions were selected to assess their efficiency using CONFIGURE. The sole criterion was to identify all possible impervious surfaces that could be made permeable. Such surfaces included roadside pavements, paths, and parking areas. From the detailed land use data extracted from Ordinance Survey (OS) MasterMap® (https://digimap.edina.ac.uk), the impervious area available for permeable interventions totalled 0.74 $km^2$, with 46% roadside pavements, 31% parking areas, and 23% paths (Figure 6a). The remaining impervious surfaces were roads and therefore unsuitable for permeable surface interventions. To assess the cost efficiency of permeable interventions at the distinct parts, the catchment was divided into zones of varying sizes, a process known as spatial discretisation (refer to Figure 6b-e). It is important to note that the term '*zone*' is used generically to represent a permeable intervention of a specific size. As CityCAT uses a high-resolution computational grid, the zones were defined as simple geometric 'boxes', with each box or zone containing many computational grid cells to represent permeable features explicitly (please refer to Figure 5a). The size of each box is determined by the area and practicality of implementing permeable interventions. Any differences in zone intervention areas will automatically be normalized by the intervention cost per unit during optimisation. The simple box-based zoning method enabled consistency between ancestor and descendant zones during spatial discretisation. Eventually, 10 larger zones (Figure 6b) were subdivided up to a total of 80 smaller zones (Figure 6e), with the majority having dimensions of approximately 340m x 180m. This scheme provided areas of appropriate size for the practical implementation of interventions. The multi-scale spatial discretization approach aimed to assess the cost-effectiveness of permeable zones at their varied sizes. The spatial discretisation scheme is further elaborated in *Supplementary Information S2 Spatial discretisation scheme*. For optimisation encoding, each permeable zone was assigned a unique ID to represent its location, with each ID serving as a decision variable. In a candidate solution, these variables can either be present (value = 1) or absent (value = 0). Therefore, the optimization of 10 permeable zones involves 10 decision variables. Similarly, 15, 40, and 80 permeable zones correspond to 15, 40, and 80 decision variables, respectively.

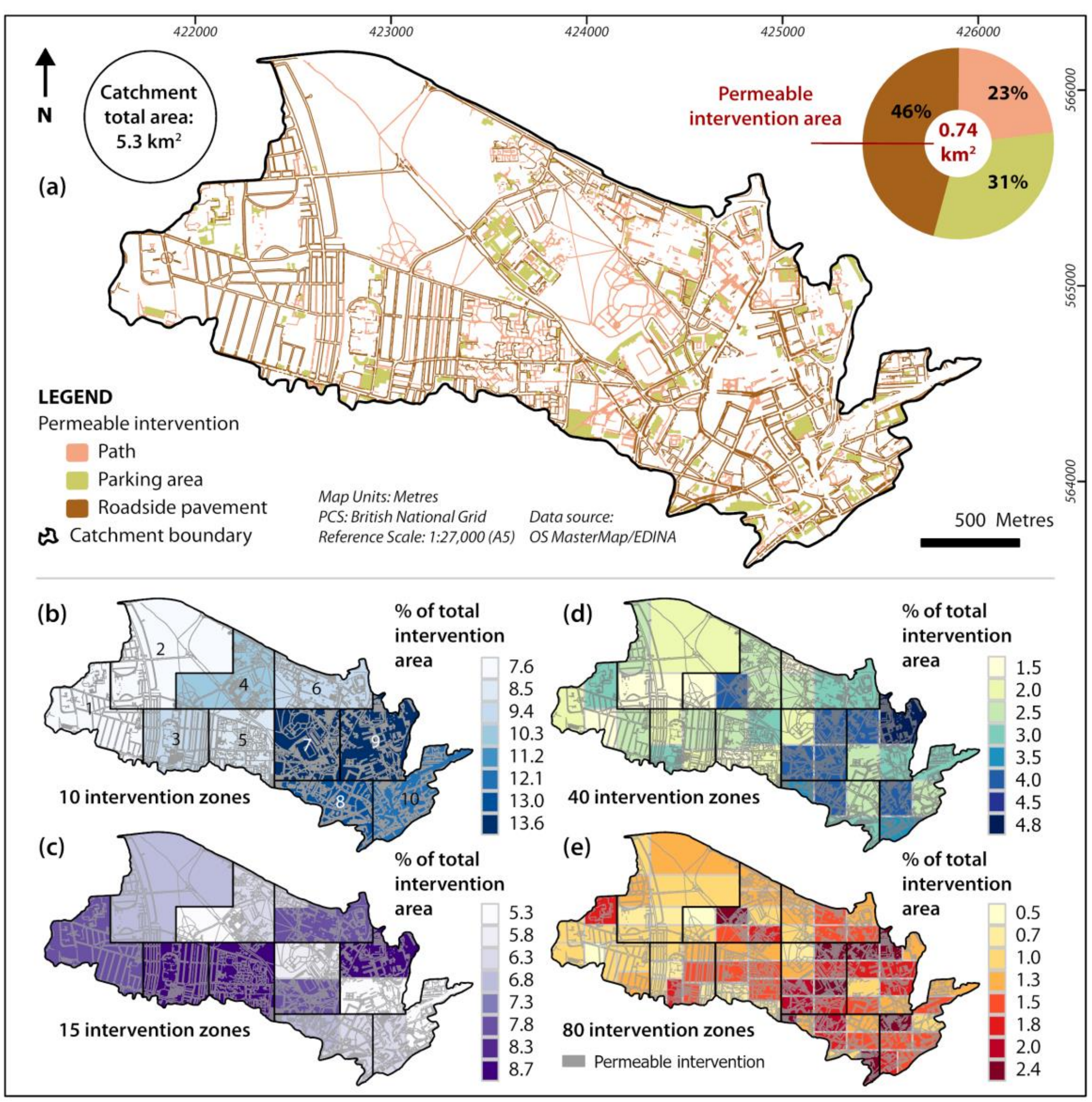

Figure 6. Map of (a) permeable surfaces and their division into (b) 10 zones, (c) 15 zones, (d) 40 zones, and (e) 80 zones. The filled colour shows the percentage of the total intervention area contained by a specific zone. 10-zone boundary (black) used as a reference for interpreting results.

### 2.3.5 CONFIGURE settings

**Cost objective function:** The unit area lifecycle cost of the permeable surface intervention was calculated using the guidelines from the Environment Agency (Gordon-Walker et al., 2007). Consumer Price Index (CPI)-based average inflation rate (2007-2022) of 2.9% was used in Equation (5) to calculate the per unit life cycle cost of permeable surface for a 40-year lifespan. This cost was then multiplied by the intervention area in each zone to obtain zone-wise lifecycle cost.

$$FV = BV(1+i)^n \qquad (5)$$

Where FV is the future value, BV is the base year value, i is the inflation rate, and n is the maintenance year. The life cycle cost of each zone was incorporated into Equation (4) to calculate the cost objective function for the optimisation.

**Risk objective function:** The exposure estimation method described in section 2.3.2 was used to proxy the risk to the buildings. From the different exposure categories, only highly exposed buildings were considered to develop the risk objective function ($F_R$) that can be expressed as follows:

$$F_R = E_B(S_z) = \sum_{i=1}^{m} B_i \times I_E \qquad z = 1,2,3....p \qquad (6)$$

$$I_E = f(d_m, d_{90th}) = \begin{cases} 1 & d_m \; AND \; d_{90th} \geq criteria \\ 0 & d_m \; OR \; d_{90th} < criteria \end{cases} \qquad (7)$$

In Equation (6) $E_B(S_z)$ represents the number of buildings highly exposed to flooding when the $z^{th}$ candidate solution is implemented. $B_i$ refers to the $i^{th}$ building within the catchment, with a total of m buildings. $I_E$ represents an exposure index, which is the function of the mean ($d_m$) and $90^{th}$ percentile ($d_{90th}$) of flood depth values surrounding $B_i$. The value of $I_E$ is set to 1 if values of $d_m$ and $d_{90th}$ around $B_i$ meet specific criteria, otherwise, the value is 0.

**NSGA-II parameters:** Table 1 presents the NSGA-II parameters and their values used in this study. The length of the candidate solution is determined by the number of permeable intervention zones. The population sizes for 10 and 15 permeable zone scenarios were chosen based on optimal solutions obtained from simulating all possible intervention options. Population size for the remaining scenarios along with the mutation type and mutation rates were selected based on NSGA-II parameter sensitivity analysis (see *Supplementary Information S5 Parameter sensitivity analysis*)

| **Parameter** | **Type** | **Parameter size/rate/value**<br>Intervention zones/decision variables<br>(possible solutions) | | | |
|---|---|---|---|---|---|
| | | 10 zones<br>($2^{10} = 1024$) | 15 zones<br>($2^{15} = 32768$) | 40 zones<br>($2^{40}$) | 80 zones<br>($2^{80}$) |
| Candidate solution length (n) | Binary matrix (number of columns) | 10 | 15 | 40 | 80 |
| Population size (p) | Binary matrix (number of rows) | 27 | 66 | 100 | 100 |
| Selection | Binary tournament | - | - | - | - |
| Cross-over | Random single-point with probability | 1.0 | 1.0 | 1.0 | 1.0 |
| Mutation | Random single bit-flip with probability | 0.4 | 0.4 | 0.4 | 0.4 |
| Stopping criteria | Maximum number of generations | 25 | 50 | 100 | 100 |

Table 1. Genetic Algorithm parameters and their values

**Integration of NSGA-II with CityCAT:** Figure S10 in *Supplementary Information S10* details the integration of NSGA-II with CityCAT. A standard directory structure for input files was established in Windows, with all CityCAT inputs fixed except for the permeable surface geometries. CityCAT parameter values were provided in the configuration files. Initially, a Python script imports the geometry files and life cycle costs for each permeable surface zone. NSGA-II then creates a population of candidate solutions. For each candidate solution, the script combines the life cycle costs and geometries of the contributing permeable zones, exports the combined geometries to a CityCAT-compatible format, and executes CityCAT. The output (maximum water depths) and building footprints, are imported back into Python to calculate the risk levels for each candidate solution. This process is repeated until all candidate solutions are evaluated for their life cycle costs and exposure levels. NSGA-II then evaluates the fitness of these solutions and performs evolutionary operations (parent selection, crossover, mutation, offspring generation, and new generation creation). This cycle continues until optimal solutions are obtained after T generations, which are then exported to a CSV file.

**Hardware specification:** CityCAT simulations were performed on a workstation with an "Intel(R) Core(TM) i9-10900X CPU @ 3.70GHz" processor and 64 GB RAM. On average, a single CityCAT simulation took a little less than a minute to complete.

# 3 Results and discussion

## 3.1 Assessing the effectiveness of CONFIGURE

### 3.1.1 Optimisation of 10-zone case

To illustrate CONFIGURE's functionality, all possible candidate solutions for 10 permeable zones were simulated to obtain the global optimal solutions for testing purposes. As shown in Figure 7, the Pareto front resulted in 27 optimal solutions. Evaluation of all options for 15 permeable zones is given in *Supplementary Information S3 Simulating all available options for 15 zones w*hile further discussion on Pareto fronts is presented in Section 3.2.

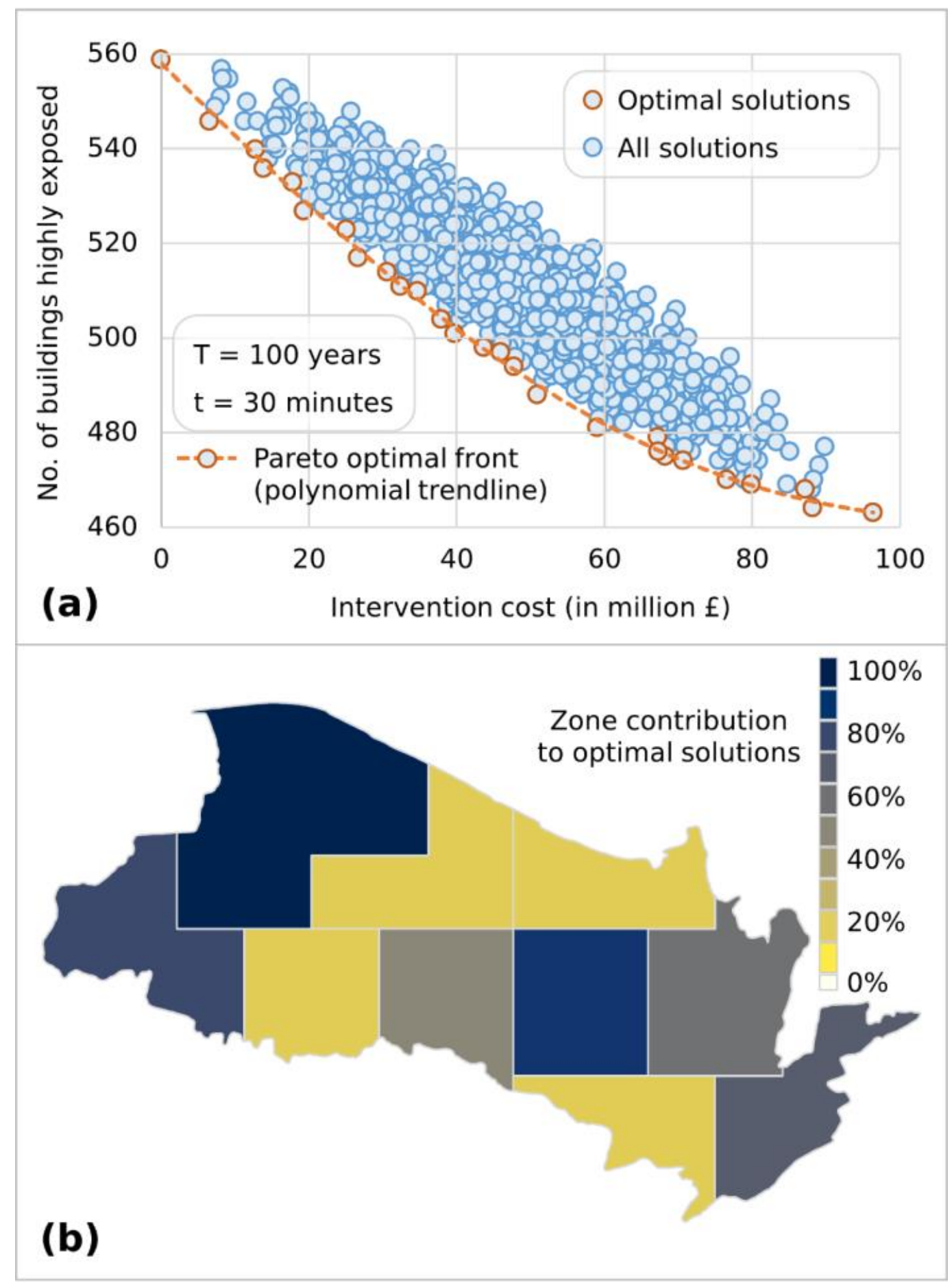

26 Figure 7. (a) All candidate solutions including optimal for 10 permeable zones, (b) contribution of

27 each permeable zone to optimal solutions

28 To assess the effectiveness of the proposed optimisation framework, a convergence test was conducted.
29 Optimal solutions obtained through simulating all possible combinations (Figure 7) were used as a
30 reference. The initial population (Figure 8a), equal to the number of reference solutions, produced the
31 first generation (Figure 8b) that brought candidate solutions nearer towards the reference Pareto front,
32 and zone-2 emerged as the highly contributing zone in generation 1. This trend continued to generation
33 2 (Figure 8c) by bringing further distinction among the zones. Generations 3-7 (Figure 8d-h) continued
34 to provide better candidate solutions, bringing them closer to the reference Pareto front and providing
35 clearer distinctions in the permeable surface zones' efficiency. Finally, in generation 8 (Figure 8i), the
36 evolving solutions completely overlapped with the reference optimal solutions, indicating convergence.
37 CONFIGURE algorithm intelligently simulated only 24% of the total candidate solutions (Figure 8j) to
38 achieve global optima, establishing a clear distinction in the zones based on their contribution to the optimal solutions. The convergence test for the 15-zone scenario is presented in S*upplementary Information S4 CONFIGURE convergence test for 15 zones*.

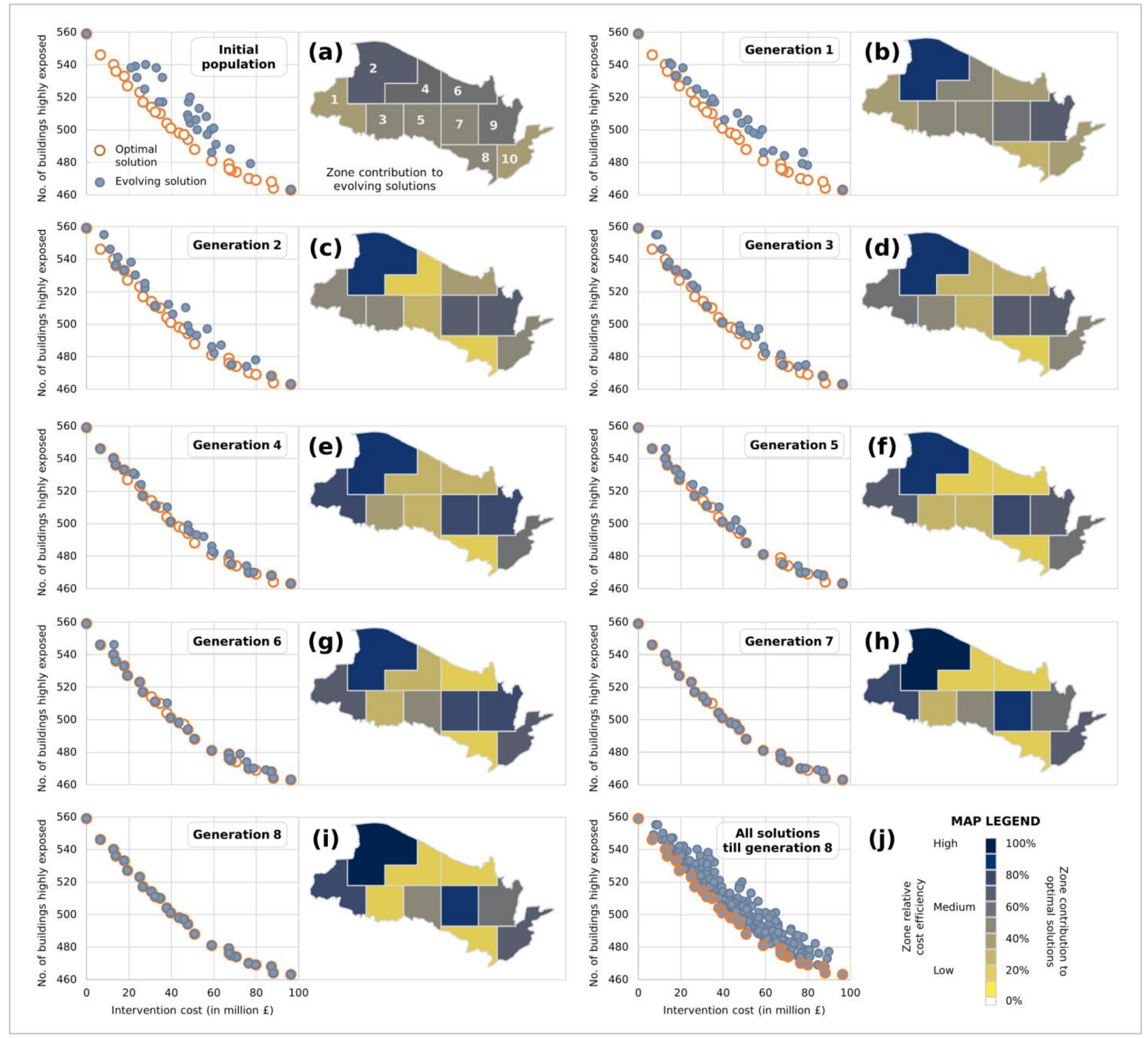

Figure 8. Convergence test for 10 permeable zones (a) initial population with reference optimal solutions, (b-i) evolving generations, and (j) total solutions tested by CONFIGURE to achieve convergence.

### 3.1.2 CONFIGURE as a rapid screening tool

The optimisation process is given in Appendix B: Animation of the optimisation process for the 40-zone scenario. As can be observed in the convergence tests in section 3.1.1 and snapshots of the 20$^{th}$ (Figure 9a) and 60$^{th}$ (optimal) generations (Figure 9b) from the animation, CONFIGURE starts

49 detecting the majority of the best and the least performing locations for permeable surface interventions
50 in the initial stages of optimisation. This demonstrates CONFIGURE's potential as a rapid screening
51 tool for examining BGI location performance.

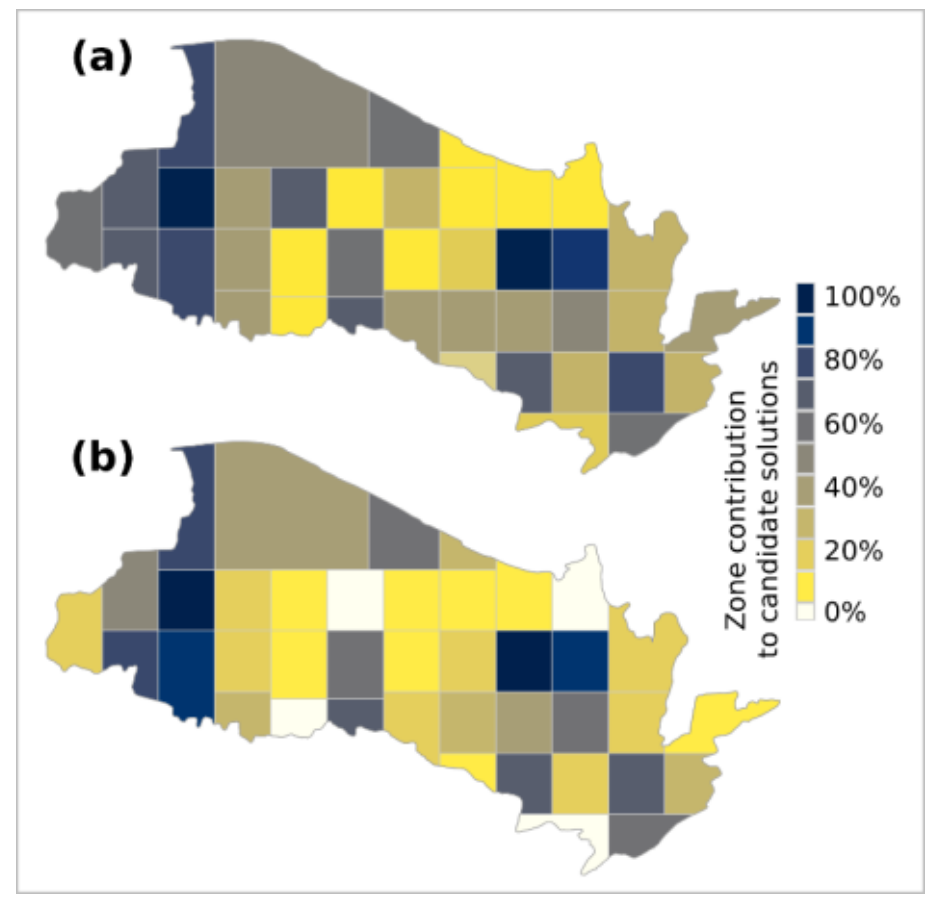

Figure 9 Zones efficiency at: (a) $20^{th}$ generation, (b) $60^{th}$ (optimal) generation during optimisation.

**3.1.3 Convergence time**

The time required to evaluate all possible solutions was calculated by multiplying the average CityCAT
simulation time by the total number of possible solutions, which are $2^n$, where n is the number of
decision variables. Figure 10 demonstrates that the time required to evaluate all possible solutions
increases exponentially as the number of zones (decision variables) grows. However, CONFIGURE
shows a behaviour change: at first, it behaves linearly, but as the number of decision variables increases
further, it starts to trend towards a straight line. *Supplementary information S6 CONFIGURE*
*convergence time* provides a detailed table on CONFIGURE's time efficiency. It is worth mentioning
that the convergence time remained the same when CityCAT was run with and without parallelisation
scheme. This is because CityCAT is already effectively parallelised and its single instance fully uses
all available processing cores and threads, leaving no room for additional parallel tasks. Running more
than one CityCAT instance in parallel divides processing resources, thereby increasing simulation time
equivalently for each of the instances.

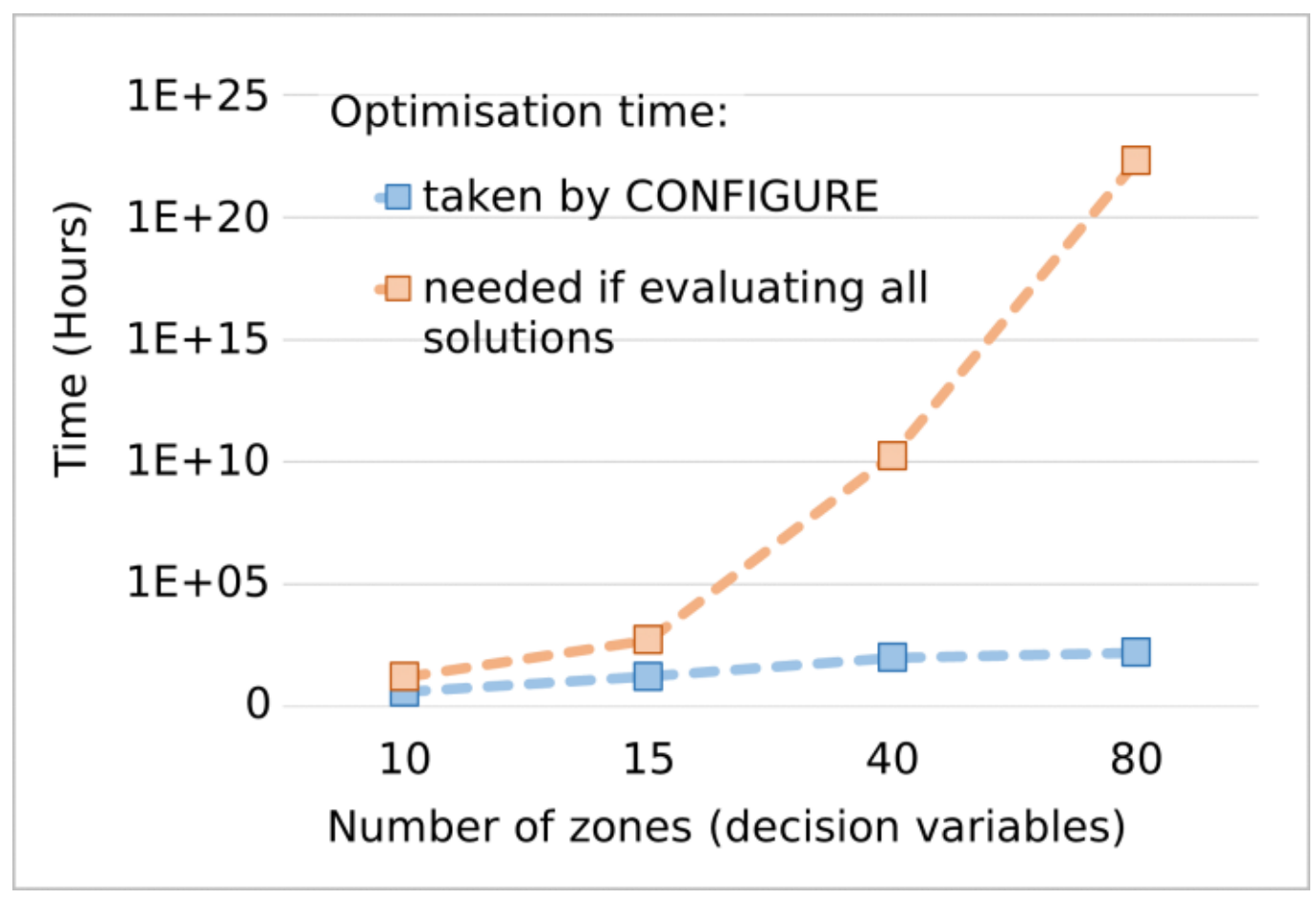

Figure 10. Time needed to evaluate all possible solutions vs. time taken by CONFIGURE to attain near-optimal solutions. The Y-axis (time) is scaled logarithmically.

## 3.2 Analysis of optimised outputs

### 3.2.1 Impact of spatial discretisation

Figure 11a shows the Pareto fronts for the four different spatially discretised scenarios, while maps in Figure 11b-e present the contribution of each permeable zone to the optimised set of solutions. It is evident from Figure 11a that the higher spatial discretisation (small zone size, more quantity) produces more and significantly better optimal solutions. In other words, the Pareto front gains more curvature towards minimal values of cost and exposure. Further, making permeable zones smaller in size offers maximum exposure reduction with reduced intervention cost (see solutions for the least exposure (463 buildings) on the Pareto fronts).

The zone contribution maps in Figure 11b-e show that irrespective of their intervention size, different zones have different cost-efficiency when they work in combination. For example, referring to Figure 11b and Figure 6b, despite having comparable intervention areas (and hence lifecycle cost), zone 7 demonstrates better cost-efficiency than zone 9 in reducing building exposure. This observation holds true for the small-zone scenarios, such as zones 712, 721 & 722 (Figure 11d & Figure 6d), and zones 7121 & 7221 (Figure 11e & Figure 6e). Moreover, despite having a comparatively smaller intervention area, zone 2 (Figure 11b & Figure 6b) contributes to over 90% of the optimal solutions, making it the most cost-effective zone. Zone 8 (Figure 11b & Figure 6b), on the other hand, performs inversely.

These results emphasize the significance of considering the location when deploying permeable features.

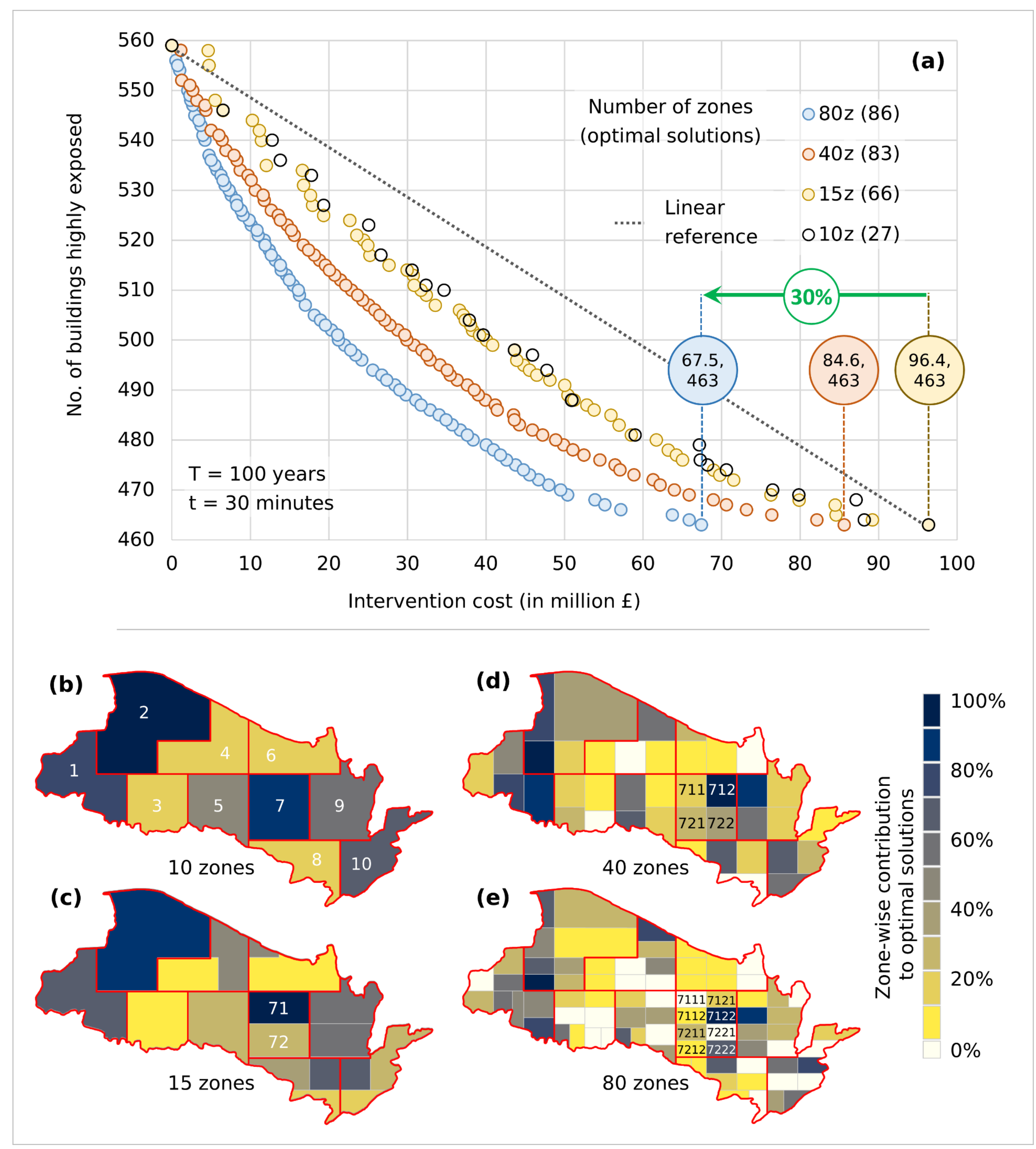

Figure 11. (a) Pareto fronts for spatially discretised scenarios. Maps depicting zone contribution to Pareto front in (b) 10-zone, (c) 15-zone, (d) 40-zone, and (e) 80-zone scenarios. The 10-zone scenario (red boundary) is used as a reference for interpreting results.

The impact of both spatial discretisation and location suitability can be better understood by further examining the zone-wise contribution maps (Figure 11b-e). For instance, when zone 7 in Figure 11b is bifurcated (Figure 11c), subzone 71 becomes the highest contributing zone, showing a better cost-effectiveness of the upper part than the lower part of zone 7. Upon further division of zone 71, subzone 712 (Figure 11d) emerges as the most cost-effective zone and so on. When comparing the boundaries of zone 7 in Figure 11b and Figure 11e, it becomes evident that only one out of its eight parts is the most economical. Two of its parts (zones 7111 and 7221) even show zero contribution to optimal solutions. Similarly, zones 1,2, 9, and 10 exhibit similar characteristics in their spatial subdivisions. Surprisingly, further divisions of less cost-effective zones 6 and 8, produced a couple of better cost-efficient parts.

The influence of zone location sensitivity and spatial discretisation on optimisation, as explained above, can be attributed to the hydrodynamics of the area. Primarily, permeable surfaces control surface run-off volume by infiltrating water into the soil subsurface. Additionally, they attenuate the velocity of flood flow by offering fractionally more friction than impervious surfaces. Based on the catchment elevation (Figure 4), the natural movement of the floodwater (Figure 12) is from the northwest (upper catchment) towards the southeast (lower catchment). Thus, despite having a smaller proportion of intervention areas, zones 1 and 2 not only reduce local exposure but also reduce flows towards other zones, indirectly contributing to reduced exposure within the territories of other zones as well. This could be the reason for the lower cost-efficiency of zone 4. A similar logic can be attributed to zone 7 and the smaller-zone scenarios. In terms of optimisation algorithm functionality, a finer-grained spatial discretisation empowers the algorithm with a range of distributed spatial options to create better cost-effective combinations.

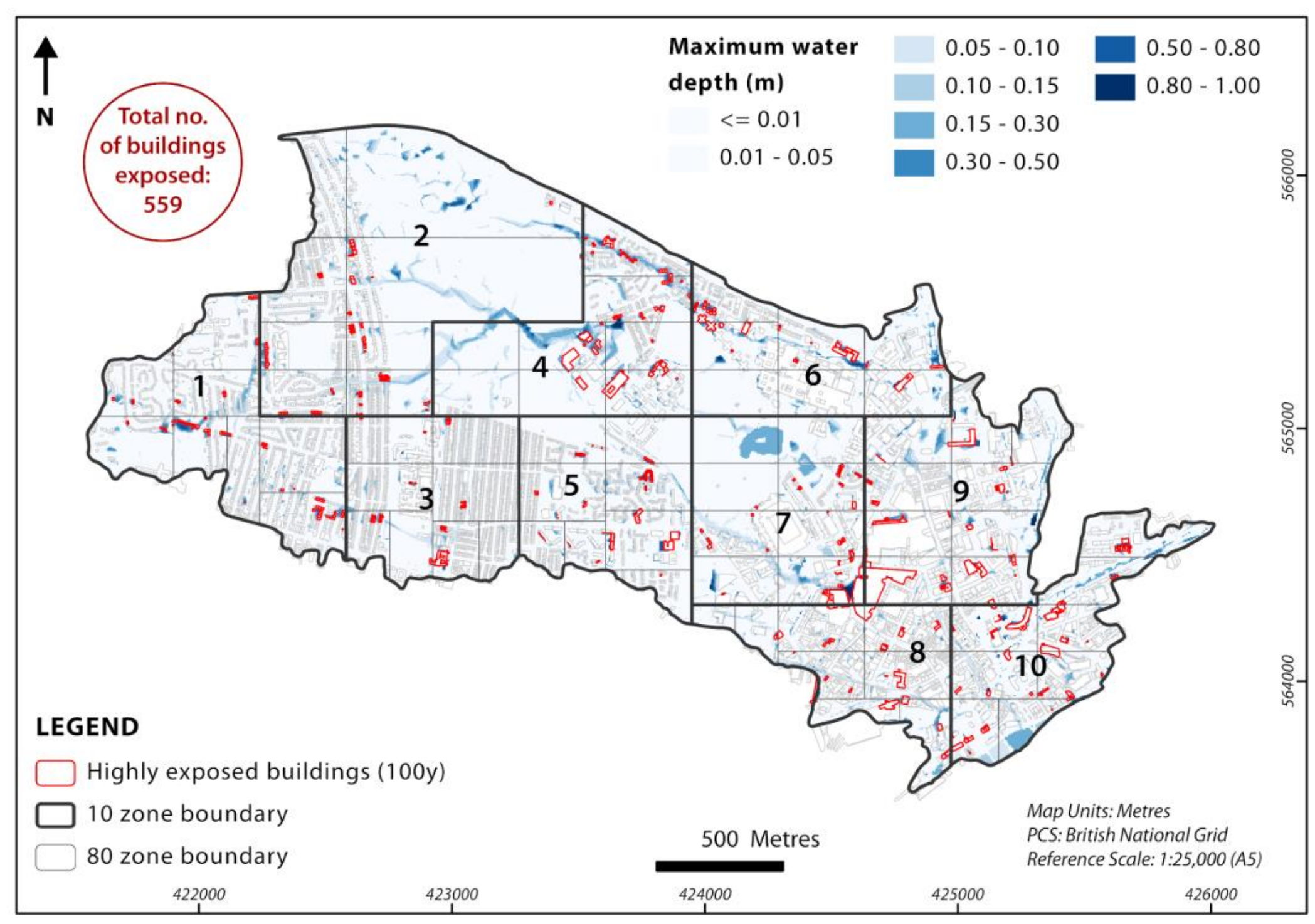

Figure 12. Exposure map for baseline scenario (no BGI intervention, return period = 100 year, duration = 30 minutes)

### 3.2.2 Cost-benefits of optimal solutions

Referring to Figure 11a and considering the best optimisation case i.e. 80-zone scenario, an initial investment up to £10 million saves approximately 40 buildings, resulting in a cost-benefit ratio of 1:4. Further investment decreases this ratio; for example, investing an additional £30 million only saves approximately 60 additional buildings, with a cost-benefit ratio of 1:2. This trend continues with additional investments. Therefore, for the current study, permeable interventions can yield a relatively better return on an investment of up to £10 million.

**Impact of different rainstorm intensities**

Figure 13 displays the optimisation outcomes of the 80-zone scenario for the 100-year and 30-year rainstorm events. The efficiency of interventions optimised for the 100-year rainstorm event was also evaluated for the 30-year return period, and vice versa. Two main results emerged are:

1. CONFIGURE performed well in optimising interventions for both rainstorm events, demonstrating its effectiveness in handling different rainstorm intensities.

2. The interventions optimised for the 30-year rainstorm event performed sub-optimally when
evaluated against the 100-year event, and vice versa. This discrepancy indicates that the location
and combinations of the most cost-efficient permeable zones may vary between different
rainstorm events, suggesting that solutions based on a single rainstorm event are not universal
(see Figure 13).

The variations in optimisation results can be attributed to the significant differences in rainfall intensity
and total rainfall amount between two distinct rainstorm events. The significant difference in rainfall
intensities probably leads to variations in exposure quantity and its distributions within the catchment
(Refer to *Supplementary Information S8 Buildings exposure to rainstorms of different return periods*),
resulting in different optimal zones and their combinations. However, it is anticipated that optimisation
results will exhibit similar characteristics for rainstorm events with smaller differences in rainfall
intensities.

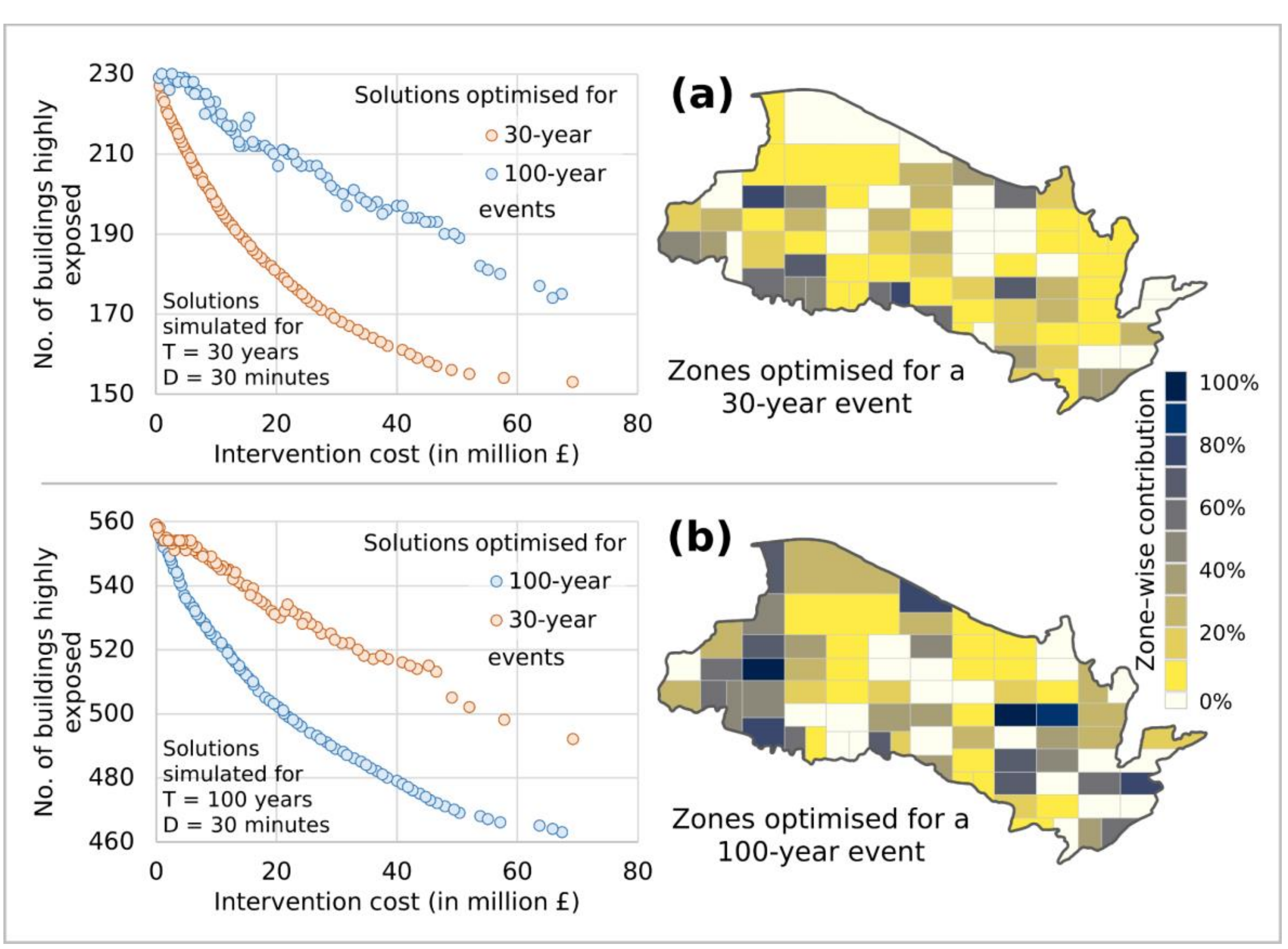

Figure 13. Optimisation of 80-zone scenario for (a) 30-year, and (b) 100-year events

### 3.3 Study limitations

There are a few limitations associated with the current study. Firstly, due to a higher computational cost, only the 2D surface module of the CityCAT was used to test the optimisation methodology in this initial study. While the addition of the storm sewer network would be expected to reduce the overall damages, especially for smaller floods, the overall spatial pattern is expected to be similar, and the study serves as a good starting point for understanding and implementing a detailed model-enabled optimisation and its related outputs. Secondly, a 5-m spatial resolution DEM was used in hydrodynamic simulations to develop and test the effectiveness of the CONFIGURE framework. However, a higher spatial resolution DEM can offer a more accurate representation of land use features, BGI interventions, and flood pathways in an urban environment and thus can influence optimal choices. Nevertheless, as with all urban flood models, a compromise must be reached between DEM grid resolution and model simulation speed. Thirdly, BGI costing was approximated by following broader guidelines provided by the UK Environment Agency (EA) in 2007. A detailed appraisal can provide more accurate costing, but the optimisation processes will remain the same. Finally, this study considered the same type of permeable surface for parking as well as pavements and paths. However, if needed, diverse types of permeable surfaces for different impermeable features can be exercised by replicating the same methodology presented in this paper.

### 3.4 Future recommendations

The results of the study reveal that the benefits of permeable interventions are relatively low, even with a higher spatial discretisation. Furthermore, an optimised solution for a single return period is not effective for others. Therefore, considering multiple return periods together for optimisation is needed. It will not only provide a robust solution to tackle climate change impacts but may also improve the return on investment. Considering other types of BGI, such as ponds or swales, can also provide a more favourable cost-benefit ratio. When aiming to enhance CONFIGURE's convergence time, if the hydrodynamic model is already parallelised or multi-thread enabled, it is advisable to consider multi-node parallelisation by configuring primary/subordinate settings instead of applying a single-machine parallelisation method. Moreover, users can also try updated variants of NSGA-II, such as Epsilon-NSGA-II, to speed up convergence time. Finally, utilisation of a fully coupled hydrodynamic model, combining a 2D surface and 1D sewer drainage modules, can evaluate surface flood risk and BGI efficiency more accurately to produce an improved urban FRM design.

## 4 Conclusions

The newly proposed Cost OptimisatioN Framework for Implementing blue-Green infrastructURE (CONFIGURE) offers BGI optimisation by simplifying the problem-framing procedure, implementing effective genetic algorithm operations, using a detailed hydrodynamic model, and introducing an effective visualisation scheme for differentiating between efficient and inefficient interventions. CONFIGURE demonstrates its capability to effectively achieve optimal solutions for various rainfall and spatial discretisation scenarios. The use of a high-resolution 2D surface flood model enables the explicit representation of permeable features, allowing for better simulation and understanding of their functions at distinct locations during different rainstorms. The time required to achieve optimal solutions depends on the number of BGI locations being optimised. Although optimising many smaller permeable surface zones takes more computation time, it results in significant cost savings. In the current study, dividing 10 large permeable zones into 80 smaller zones allowed CONFIGURE to save approximately 30% of the investment by eliminating ineffective permeable areas. However, the efficiency of permeable zones optimised for higher rainfall intensities does not translate well to lower rainfall intensities, and vice versa. Therefore, selecting an appropriate rainstorm return period for BGI location optimization is critical. Regardless of the return period chosen, it is preferable to implement many small BGI interventions across the catchment to achieve the best cost-benefit ratio. CONFIGURE's ability to maximise investment efficiency using an explicit flood model makes it a promising tool for designing BGI for sustainable flood risk management in urban areas.

## Funding

This research work has been carried out under the ONE Planet Doctoral Training Partnership, funded by the Natural Environmental Research Council (NERC) through grant NE/S007512/1.

## Acknowledgements

The authors would like to thank Chris Iliadis of Newcastle University for providing an updated Python script for the building-level exposure analysis tool.

## Conflict of interest

The authors declare no conflict of interest.

## Appendix A: Supplementary information

Attached: *Supplementary information.docx*

## Appendix B: Animation of the optimisation process

Attached: *Animation of optimisation process.mp4*

## Appendix C: Availability of Python code

Repository name: CONFIGURE v1.0; Code access: *https://github.com/asidurrehman/configure10*; Developer: Asid Ur Rehman (asid-ur-rehman2@newcastle.ac.uk); Year first published: 2024 (expected); Hardware required: PC; Operating system: Windows, Linux, iOS; Software required: Anaconda 2.3.1 & Spyder 5.4.2; Programming language: Python 3.9; Python packages required: Numpy 1.26.0, Pandas 2.1.1, Matplotlib 3.8.0; Availability and cost: Open source; Licence: Apache 2.0